\documentclass{article}

\PassOptionsToPackage{numbers, compress}{natbib} % <--- ADD THIS

\usepackage[final,main]{neurips_2026}

\usepackage[utf8]{inputenc} % allow utf-8 input
\usepackage[T1]{fontenc}    % use 8-bit T1 fonts
\usepackage{hyperref}       % hyperlinks
\usepackage{url}            % simple URL typesetting
\usepackage{booktabs}       % professional-quality tables
\usepackage{amsfonts}       % blackboard math symbols
\usepackage{amsmath}        % math typesetting
\usepackage{amssymb}        % symbols
\usepackage{nicefrac}       % compact symbols for 1/2, etc.
\usepackage{microtype}      % microtypography
\usepackage{xcolor}         % colors
\usepackage{xspace}         % spacing
\usepackage{cleveref}       % referencing and linking
\usepackage{siunitx}        % scientific units
\usepackage[normalem]{ulem}
\usepackage{graphicx}
\usepackage[labelfont=bf]{caption}
\usepackage{tcolorbox}
\usepackage{stfloats}
\fnbelowfloat
\usepackage{wasysym}
\usepackage{wrapfig}

\tcbuselibrary{skins}

\usepackage{etoolbox}

\def\importantboxstyle{rounded}

\ifdefstring{\importantboxstyle}{modern}{
    \newtcolorbox{importantbox}{
        enhanced,
        colback=gray!10,
        frame hidden,
        borderline west={2pt}{0pt}{black!60},
        sharp corners,
        boxsep=2pt,
        left=6pt
    }
}{}

\ifdefstring{\importantboxstyle}{outline}{
    \newtcolorbox{importantbox}{
        colback=white,
        colframe=black,
        boxrule=0.5pt,
        arc=0pt,
        left=3mm,
        right=3mm,
        top=2mm,
        bottom=2mm
    }
}{}

\ifdefstring{\importantboxstyle}{rounded}{
    \newtcolorbox{importantbox}{
        colback=black!5,
        colframe=black!5,
        boxrule=0pt,
        arc=2mm,
        left=2.5mm,
        right=2.5mm,
        top=1.8mm,
        bottom=1.8mm
    }
}{}

\newcount\eqtagmode
\usepackage{tikz}
\usetikzlibrary{tikzmark, fit, backgrounds}
\usepackage{soul}

\makeatletter
\newcommand{\tagequations}{\@ifstar{\tagequations@text}{\tagequations@auto}}

\newcommand{\tagequations@auto}[3][]{%
    \ifcase\eqtagmode
        #3%
    \or
        \ifmmode
            \tikz[baseline, remember picture]{
                \node[anchor=base, inner xsep=4pt, inner ysep=2pt, rounded corners=4pt, fill=#2, #1]
                {$\displaystyle #3$};%
            }
        \else
            \smash{\tikz[baseline, remember picture]{%
                \node[anchor=base, inner xsep=4pt, inner ysep=1.2pt, rounded corners=4pt, fill=#2, #1]
                {$\displaystyle #3$};%
            }}
        \fi
    \or
        \ifmmode
            \tikz[baseline, remember picture, inner sep=0pt, outer sep=0pt]{%
                \node[anchor=base, inner sep=0pt, outer sep=0pt, trim left=0pt, trim right=0pt, #1] (hl) {$\displaystyle #3$};%
                \begin{pgfonlayer}{background}%
                    \draw[line width=3pt, color=#2]([yshift=1pt]hl.south west) -- ([yshift=1pt]hl.south east);%
                \end{pgfonlayer}%
            }
        \else
            {
                \setulcolor{#2}%
                \setul{0pt}{3pt}%
                \smash{\ul{\xspace#3}}%
            }
        \fi
    \fi
}

\newcommand{\tagequations@text}[3][]{%
    \ifcase\eqtagmode
        #3
    \or
        \smash{\tikz[baseline, remember picture]{%
            \node[anchor=base, inner xsep=4pt, inner ysep=1.2pt, rounded corners=4pt, fill=#2, #1]
            {$\displaystyle #3$};%
        }}
    \or
        {
            \setulcolor{#2}%
            \setul{0pt}{3pt}%
            \smash{\ul{\xspace#3}}%
        }
    \fi
}

\definecolor{color_decoder}{RGB}{214,239,224} 
\definecolor{color_bio}{RGB}{235,210,250}
\definecolor{color_encoder}{RGB}{180,200,255}
\definecolor{color_feedback}{RGB}{255,225,190}
\definecolor{color_adaptation}{RGB}{180,190,200}
\definecolor{color_cl}{RGB}{50, 177, 102}
\definecolor{color_hri}{RGB}{162,86,219}
\definecolor{color_equal}{RGB}{0,0,0}

\DeclareRobustCommand{\tagEncoder}{\@ifstar{\tagequations@text{color_encoder!70}}{\tagequations@auto{color_encoder!70}}}
\DeclareRobustCommand{\tagDecoder}{\@ifstar{\tagequations@text{color_decoder!80}}{\tagequations@auto{color_decoder!80}}}
\DeclareRobustCommand{\tagBio}{\@ifstar{\tagequations@text{color_bio!80}}{\tagequations@auto{color_bio!80}}}
\DeclareRobustCommand{\tagFeedback}{\@ifstar{\tagequations@text{color_feedback!80}}{\tagequations@auto{color_feedback!80}}}
\DeclareRobustCommand{\tagAdaptation}{\@ifstar{\tagequations@text{color_adaptation!30}}{\tagequations@auto{color_adaptation!30}}}
\makeatother

\newcommand{\affcl}{\textcolor{color_cl}{\scalebox{1.0}{$\LEFTcircle$}}}
\newcommand{\affhri}{\textcolor{color_hri}{\scalebox{1.0}{$\RIGHTcircle$}}} 
\newcommand{\equalCont}{\textcolor{color_equal!40}{\scalebox{1.0}{$\CIRCLE$}}}

\title{Embodied Neurocomputation:\\A Framework for Interfacing Biological Neural Cultures with Scaled Task-Driven Validation}

\author{
Johnson Zhou\textsuperscript{\affcl\equalCont} \quad Daniel Tanneberg\textsuperscript{\affhri\equalCont} \\
\textbf{Forough Habibollahi}\textsuperscript{\affcl} \quad \textbf{Alon Loeffler}\textsuperscript{\affcl} \quad \textbf{Kiaran Lawson}\textsuperscript{\affcl} \quad \textbf{Valentina Baccetti}\textsuperscript{\affcl}\\
\textbf{Kwaku Dad Abu-Bonsrah}\textsuperscript{\affcl} \quad \textbf{Candice Desouza}\textsuperscript{\affcl} \quad \textbf{Finn Doensen}\textsuperscript{\affcl} \quad \textbf{Bradley Watmuff}\textsuperscript{\affcl}\\
\textbf{Daria Kornienko}\textsuperscript{\affcl} \quad \textbf{Azin Azadi}\textsuperscript{\affcl} \quad \textbf{Justin L. Bourke}\textsuperscript{\affcl}\\
\textbf{Bernhard Sendhoff}\textsuperscript{\affhri} \quad \textbf{Brett J. Kagan}\textsuperscript{\affcl}
\vspace{0.2cm}
\\
\textsuperscript{\affcl}Cortical Labs, Australia \quad \textsuperscript{\affhri}Honda Research Institute Europe, Germany \quad \textsuperscript{\equalCont}equal contribution
}

\begin{document}

\maketitle

\begin{abstract}
Biological neural networks (BNNs) have been established as a powerful and adaptive substrate that offer the potential for incredibly energy and data efficient information processing with distinct learning mechanisms. Yet a core challenge to utilizing BNN for neurocomputation is determining the optimal encoding and decoding mechanisms between the traditional silicon computing interface and the living biology. Here, we propose an Embodied Neurocomputation framework as a systems-level approach to this multi-variable optimization encoding/decoding problem. We operationalize this approach through the first large-scale parameter optimization of encoding configurations for a BNN agent performing closed-loop navigation along an odor-style gradient in a simulated grid-world. Despite the relative simplicity of the task, the biological interactions gave rise to a massive multi-combinatorial search space for optimal parameters. By considering how the components of the system are interconnected and parameterized, we evaluated approximately 1,300 parameter combinations, over 4,000 hours of real-time agent-environment interactions, to identify 12 configurations that consistently demonstrated learning across multiple episodes. These configurations achieved significantly higher task performances than optimized silicon-based DQN agents under the same interaction budget. These findings represent an initial step toward robust and scalable goal-oriented learning using BNNs. Our framework establishes a foundation for applying task-driven neurocomputing and supports the development of field-wide benchmarks. In the long term, this work supports the development of hybrid bio-silicon architectures capable of efficient, adaptive and real-time computation, including the potential for robotic control applications.
\end{abstract}

% Units of measure
\newcommand{\uAmp}{\si{\micro\ampere}\xspace}
\newcommand{\uSec}{\si{\micro\second}\xspace}
\newcommand{\uVolt}{\si{\micro\volt}\xspace}
\newcommand{\uM}{\si{\micro\text{M}}\xspace}
\renewcommand{\uL}{\si{\micro\litre}\xspace}
\renewcommand{\ug}{\si{\micro\gram}\xspace}

% Common
\newcommand{\Real}{\mathbb{R}}
\newcommand{\Binary}{\{0,1\}}

% Formula
\newcommand{\Cdot}{\,\cdot\,}
\newcommand{\BNN}{b}
\newcommand{\BNNAdaptation}{g}
\newcommand{\Timesteps}{t}
\newcommand{\Input}{\mathbf{x}_{\Timesteps}}
\newcommand{\Output}{\mathbf{y}_{\Timesteps}}
\newcommand{\AdaptationInput}{\mathbf{i}}
\newcommand{\StimMatrix}{\mathbf{u}_{\Timesteps}}
\newcommand{\StimMatrixFeedback}{\Bar{\mathbf{u}}_{\Timesteps}}
\newcommand{\ResponseMatrix}[1][\relax]{
  \ifx\relax#1\relax
    \mathbf{v}_{\Timesteps}
  \else
    \mathbf{v}_{#1}
  \fi
}
\newcommand{\ResponseSpikeMatrix}{\ResponseMatrix'}
\newcommand{\SpikeDetection}{d'}
\newcommand{\Encoder}{e}
\newcommand{\Decoder}{d}
\newcommand{\NeuroComputing}{f}

\newcommand{\Channels}{C}
\newcommand{\InputTime}{\tau_{\text{in}}}
\newcommand{\OutputTime}{\tau_{\text{out}}}

\newcommand{\Frequency}{F}
\newcommand{\FrequencyMin}{\Frequency_{\text{min}}}
\newcommand{\FrequencyMax}{\Frequency_{\text{max}}}

\newcommand{\Current}{I}
\newcommand{\PulseWidth}{w}

\newcommand{\ParametersAll}[1][\relax]{\ifx\relax#1\relax\theta\else\theta_{#1}\fi}
\newcommand{\ParametersEncoder}[1][\relax]{\ifx\relax#1\relax\theta_{\Encoder}\else\theta_{\Encoder,#1}\fi}
\newcommand{\ParametersBNN}[1][\relax]{\ifx\relax#1\relax\theta_{\BNN}\else\theta_{\BNN,#1}\fi}
\newcommand{\ParametersBNNNext}[1][\relax]{\ifx\relax#1\relax\theta^{'}_{\BNN}\else\theta^{'}_{\BNN,#1}\fi}
\newcommand{\ParametersDecoder}[1][\relax]{\ifx\relax#1\relax\theta_{\Decoder}\else\theta_{\Decoder,#1}\fi}
\newcommand{\ParametersFeedback}[1][\relax]{\ifx\relax#1\relax\theta_{\Feedback}\else\theta_{\Feedback,#1}\fi}

\newcommand{\ParametersStim}{\ParametersAll_{\text{stim}}}
\newcommand{\ParametersTask}{\ParametersAll_{\text{task}}}

\newcommand{\ParamFrequencyMin}{\theta_{\Encoder1}}
\newcommand{\ParamFrequencyMax}{\theta_{\Encoder2}}
\newcommand{\ParamStimAmp}{\theta_{\Encoder3}}
\newcommand{\ParamStimPulseWidth}{\theta_{\Encoder4}}
\newcommand{\ParamTickRate}{\theta_{\Encoder5}}
\newcommand{\ParamTicksPerStep}{\theta_{\Encoder6}}

\newcommand{\Metric}{\text{Metric}}
\newcommand{\Score}[1][\relax]{\ifx\relax#1\relax\text{Score}\else\text{Score}_{#1}\fi}

\newcommand{\Feedback}{r}
\newcommand{\Reward}{\Feedback^{+}}
\newcommand{\Punishment}{\Feedback^{-}}

\newcommand{\ScalarInput}{x}
\newcommand{\ScalarInputMin}{\ScalarInput_{\text{min}}}
\newcommand{\ScalarInputMax}{\ScalarInput_{\text{max}}}

\newcommand{\Celcius}{$^\circ\text{C}$}

\section{Introduction}
\label{section:introduction} 
The rapid growth of artificial intelligence and machine learning has pushed computational demands to unprecedented levels, exposing the limitations of traditional silicon-based architectures. As conventional hardware nears the physical boundaries of Moore's Law and struggles with the energy inefficiencies of the von Neumann bottleneck, the ecological and economic costs of scaling current computing paradigms have become unsustainable \cite{strubell_energy_2019,mehonic_brain_2022}. In contrast, biological neural networks (BNNs) perform highly complex, non-linear processing, including continuous learning, robust pattern recognition, and adaptive decision-making, while using only a fraction of the power required by digital systems \cite{bullmore_economy_2012,Smirnova_2023_Organoid,Wegmeth_2025_Green}. Bridging this efficiency gap has motivated the search for fundamentally new computational substrates. By harnessing the intrinsic processing power of living BNN cultures, \textit{neurocomputing} represents a critical paradigm shift: moving away from energy-hungry silicon toward engineered embodied biological systems capable of inherently efficient, intelligent computation.

Despite neurocomputing's promise, a fundamental bottleneck in functional \textit{in vitro} systems lies at the interface between the digital environment and biological substrate. Micro-Electrode Arrays (MEAs) are the primary platform for this bidirectional communication, enabling both recording from and stimulation of cultured neural networks \cite{wagenaar_controlling_2005, poli_functional_2015}. However, while \textit{decoding} network activity has advanced substantially~\cite{cunningham2014dimensionality, gallego2017neural}, the inverse challenge, \textit{encoding} meaningful digital information into a biological format, remains profoundly difficult~\cite{shahaf_learning_2001}. Unlike deterministic silicon pathways, biological networks are adaptive, noisy, and non-stationary \cite{valverde_breakdown_2017}. As a result, translating structural data into electrical stimuli often yields unpredictable network states, as static or poorly tuned encoding schemes struggle to reliably drive the non-linear dynamics required for computation \cite{moriya_quantitative_2019}.

Current efforts to drive information processing in MEA-embodied networks rely heavily on heuristic or ad-hoc stimulation protocols. Although studies have demonstrated closed-loop control, pattern recognition, and task-specific learning \cite{wagenaar_controlling_2005, alam_el_din_human_2025,kagan_neurons_2022,maurer2026reinforcement,Loeffler_2026_Handwritten}, they typically use highly constrained encoding parameters. Consequently, systematic literature exploring the multidimensional parameter space needed for effective information translation remains limited. Biological encoding spans a vast stimulation space, including frequency, amplitude, pulse width, waveform morphology, and spatiotemporal distribution~\cite{tanveer_starting_2025}. Without broad evaluation across these parameters, it remains unclear which encoding regimes optimally couple with network biophysics to produce robust plasticity and learning \cite{alam_el_din_human_2025}.

While methods for culturing and assembling BNNs continue to evolve (e.g.,~\cite{rosebrock_enhanced_2022, abu-bonsrah_novel_2026}), the engineering principles needed to harness their computational capacity remain underexplored. This capacity emerges from the structural and temporal interplay between external information forces and internal physiological states~\cite{kagan_neurons_2022,kagan_two_2025}, requiring systematic interface optimization. To address this gap, this study moves beyond isolated proof-of-concept stimulation paradigms.

\begin{figure}[t]
    \centering
    \includegraphics[width=1.0\linewidth]{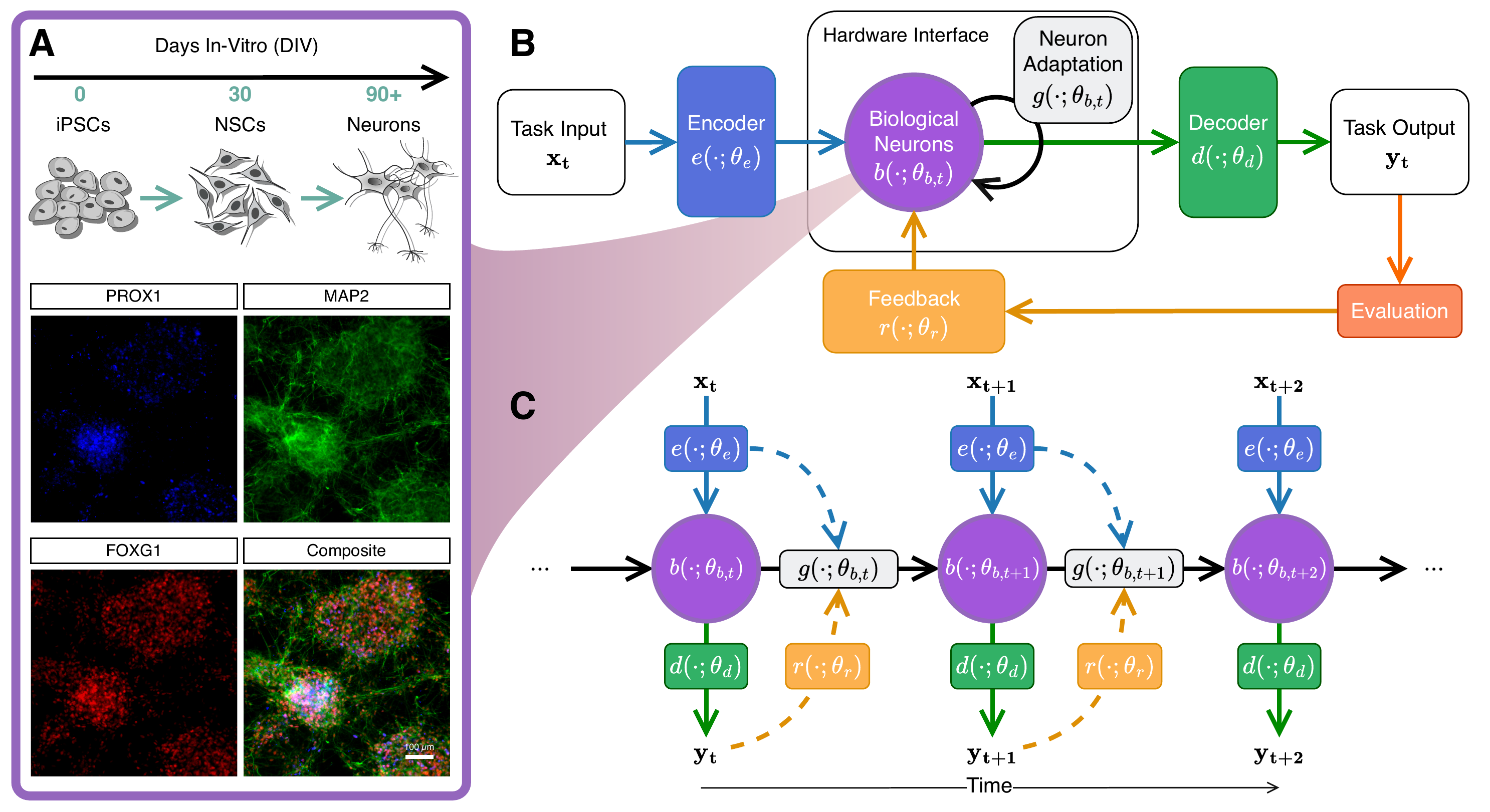}
    \caption{
    Schematic of the proposed Embodied Neurocomputation Framework defined in \Cref{section:definition}. 
    \textbf{A}: Preparation and characterization of the biological neural substrate. 
    (\text{\textit{Top}}) Timeline of neuronal differentiation from induced Pluripotent Stem Cells (iPSCs) via Neural Stem Cells (NSCs) to mature neurons at 90 days \textit{in vitro} (DIV) 
    (\text{\textit{Bottom}}) Immunocytochemical characterization of neuronal identity. Micrographs show distinct labeling for PROX1, MAP2, and FOXG1, confirming the cortical and hippocampal makeup of the computational network.
    \textbf{B}: Functional mapping of framework components involved in transforming inputs to outputs ($\Input \mapsto \Output$) at timestep $t$ through a silicon-biological hardware interface.
    \textbf{C}: Evolving temporal dynamics driven by biological neuron adaptation $\tagAdaptation*{\BNNAdaptation(\Cdot;\Cdot)}$ as it is influenced by encoding $\tagEncoder*{\Encoder(\Cdot;\Cdot)}$ and feedback $\tagFeedback*{\Feedback(\Cdot;\Cdot)}$ mechanisms.
    }
    \vspace{-10pt}
    \label{fig:framework_schematic}
\end{figure}

\textbf{In this study}, we first introduce a formal \textit{Embodied Neurocomputation Framework} (\Cref{section:definition}), which conceptualizes the interface between digital and biological substrates as a multi-variable optimization problem composed of four interdependent modules: (1) encoding of task information into electrical stimuli, (2) biological transformation via the neural network's intrinsic dynamics, (3) decoding of neural responses into task-relevant outputs, and (4) feedback to drive adaptation. 

Secondly, to validate this framework, we conduct a large-scale empirical evaluation (\Cref{section:empirical}) within a simulated goal-driven navigation environment. By integrating 26 individual BNN cultures via the Cortical Labs CL1 platform~\cite{kagan_cl1_2025} with an automated hyperparameter optimization pipeline, we systematically screened approximately 1,300 encoding configurations over 4,000 hours of real-time agent-environment interactions.\footnote{The dataset and experiment code will be released soon.} We find that: (1) specific parameter subsets enable robust learning across multiple episodes, and (2) optimized biological agents can significantly outperform silicon-based Deep Q-Networks, which serves as a standardized benchmark for discrete action-space interactions, when evaluated over an equivalent number of training steps. To our knowledge, this is the first study to optimize neurocomputing parameters at scale, establishing a systematic and data-driven foundation for the future of information processing using biological neural substrates.

\section{Embodied Neurocomputation Framework}\label{section:definition}

We define \textit{Embodied Neurocomputation} as a computational framework that interfaces biological neural networks (BNNs) with conventional computers. While physical implementations utilize specific hardware such as multi-electrode arrays (MEAs) that mediates BNN interaction in the context of electrical activity and stimulus, the framework remains hardware-agnostic by encapsulating these physical interactions within defined mapping functions. 

We approach this framework through systems-level thinking since each of its components are interconnected and highly parameterized. Adjusting any single part of the system changes how the entire system responds. Consequently, we treat the configuration of a neurocomputational task as a multi-variable optimization problem, where the objective is to find the optimal set of transformations across all parameters to achieve a specific computational goal. This approach distinguishes our work from existing definitions of Neural Computers~\cite{Zhuge_2026_Neural,tanneberg2020evolutionary} which focus on internalizing computational primitives to a latent state of artificial neural networks. In this section, we introduce the key elements of the neurocomputation framework in an abstract and generally applicable manner. In \Cref{section:empirical}, we concretely apply this framework and use it to optimize task oriented parameters at scale.

We model neurocomputation $\NeuroComputing(\Cdot;\ParametersAll[t])$ at time $t$, with parameter set $\ParametersAll[t]$, as sequential feed-forward mappings that transform elements of an input sequence ($\Input$) into corresponding elements of an output sequence ($\Output$) at timestep ($t$) (\Cref{fig:framework_schematic}):
%
% Long version
%
% \begin{align}
%     \Output &= \NeuroComputing(\theta, \Input), \; \text{comprising of} \\
%     \StimMatrix &= \Encoder(\ParametersEncoder, \Input), \\
%     \ResponseMatrix &= \BNN(\ParametersBNN, \StimMatrix),  \\
%     \Output &= \Decoder(\ParametersDecoder, \ResponseMatrix), \\
%     \Score &= \Metric(\Output)
% \end{align}
% 

%
% Short version
\vspace{-0.6cm}
\begin{gather}
    \Output
=
    f(\Input;\ParametersAll[t])
= 
    \left[
        \tagDecoder{\Decoder(\Cdot ; \ParametersDecoder)}  \circ \tagBio{\BNN(\Cdot ; \ParametersBNN[t])} \circ \tagEncoder{\Encoder(\Cdot ; \ParametersEncoder)}
    \right](\Input), \, \text{composed of}
    \label{eq:model_composition}\\
    \StimMatrix = \tagEncoder{\Encoder(\Input ; \ParametersEncoder)}, \; \ResponseMatrix = \tagBio{\BNN(\StimMatrix ; \ParametersBNN[t] )}, \; \Output = \tagDecoder{\Decoder(\ResponseMatrix ; \ParametersDecoder )}, \, \text{evaluated by}\\
    \Score = \Metric(\Output) ~,
\end{gather}
where $\Input$ is the initial input space of information, $\Output$ is the final task-relevant output space and $\Encoder(\Cdot ; \ParametersEncoder)$, $\BNN(\Cdot ; \ParametersBNN[t])$, and $\Decoder(\Cdot ; \ParametersDecoder)$ represent the sequential \tagEncoder{\textit{encoding}}, \tagBio{\textit{biological transformation}}, and \tagDecoder{\textit{decoding}} functions respectively. The output sequence can be evaluated by a $\Metric(\cdot)$ that produces a \textit{task score} that is indicative of the performance as it relates to its intended task outcomes. We separately define a mechanism for driving BNN adaptation towards high-level objectives through \tagFeedback{\textit{feedback}} based on measurable performance characteristics. The feedback function $\Feedback(\Score ; \ParametersFeedback)$, is parameterized by $\ParametersFeedback$, guided by the score produced by the defined metric:
\vspace{-0.1cm}
\begin{equation}
    % \Feedback: \ParametersBNN[t] \mapsto \ParametersBNNNext[t] \quad \text{where} \quad \BNN(\StimMatrix; \ParametersBNN[t]) \neq \BNN(\StimMatrix; \ParametersBNNNext[t])
    \ParametersBNN[t+1] = \tagAdaptation{\BNNAdaptation(\tagFeedback{\Feedback(\Score ; \ParametersFeedback)}; \ParametersBNN[t])}, \quad \text{where} \; \BNN(\StimMatrix; \ParametersBNN[t]) \neq \BNN(\StimMatrix; \ParametersBNN[t+1])~.
\end{equation}
A key differentiator of neurocomputation relative to conventional computing is that the core transformation is driven by biology, defined here as the mapping $\BNN(\Cdot;\Cdot)$, which is inherently stochastic and adaptive as modeled by the \tagAdaptation{\textit{biological adaptation}} function $\BNNAdaptation(\Cdot;\Cdot)$ (see~\Cref{sec:biological_transform}). Conceptually, encoding and decoding serve as interfaces to modulate the inputs and outputs of a biological system whose underlying biophysical dynamics are significantly more complex and less explicitly modeled than the algorithmic structures used in silicon-based machine learning. Building on this, we differentiate \textit{learning} as the process of driving BNN adaptations to improve measurable task-oriented performance, from \textit{training}, which refers to the algorithmic optimization used in artificial neural networks. This terminology emphasizes that biological adaptation emerges from intrinsic biophysical plasticity rather than gradient-based parameter updates. 

Designing a neurocomputation application requires the specification of each functional component and its respective parameters. We define the total parameter space at timestep $t$ as $\ParametersAll[t] = \ParametersEncoder \cup \ParametersBNN[t] \cup \ParametersDecoder \cup \ParametersFeedback$, comprising both tunable hyperparameters (\textit{training}) ($\ParametersEncoder, \ParametersDecoder, \ParametersFeedback$) and adaptive biological parameters (\textit{learning}) ($\ParametersBNN$). 
This unifies common BNN application paradigms, from physical reservoirs~\cite{cai_brain_2023,sumi_biological_2023} to goal-driven objectives via purposefully designed or optimized feedback~\cite{kagan_neurons_2022,maurer2026reinforcement,robbins_goal-directed_2026,khajehnejad_dynamic_2025,habibollahi_biological_2022}, enabling systematic optimization across the neurocomputational pipeline. Selecting an optimal configuration in this time-dependent high-dimensional space is non-trivial, requiring simultaneous optimization of design variables and evolving biological dynamics toward a specific task objective.

\subsection{Encoding and Decoding}\label{section:define_encoding}

We define \textit{encoding} as the process of transforming task specific information into a format that can be communicated to BNNs while maintaining compatibility with hardware responsible for biological-digital interface. While our framework is agnostic to specific types of hardware, we formulate encoding concretely in the context of electrical stimulation using micro-electrode arrays (MEAs), which is currently one of the most common forms of BNN interfacing technology \cite{khajehnejad2024graph,alam_el_din_human_2025,Hua_2025_Microelectrode}. Encoder functions based on MEA technology can be modeled as a transformation of task specific information into a stimulation matrix, subsequently delivered to the BNNs as electrical stimulation. We specify the \textit{stimulation matrix} ($\StimMatrix \in \Binary^{\Channels \times \InputTime}$) as a spatio-temporal representation of the encoding stimulus. This matrix encapsulates the physical parameters of stimulation into a single computation structure, where $\Channels$ is the stimulation channels and $\InputTime$ is the total number of pulse delivery time steps. 

The encoder is parameterized by the union of two distinct sets, $\ParametersEncoder = \ParametersTask \cup \ParametersStim$, reflecting its dual responsibility in translating task-specific requirements and configuring hardware-interfacing protocols. While \textit{task parameters} ($\ParametersTask$) are often arbitrary and depend on the specific computational objective, interface parameters ($\ParametersStim$) govern the physical delivery of stimulation, including but not limited to, stimulation frequency, stimulation amplitude, and pulse width. We provide concrete examples in \Cref{section:empirical} as part of our empirical evaluation. 

\textit{Decoding} is the inverse process of encoding, where we transform BNN responses captured by interfacing hardware into task-relevant formats. In MEA-based systems, we model the decoder as a function that transforms a response matrix ($\ResponseMatrix \in \Real^{\Channels \times \OutputTime}$), which captures continuous real-valued electrical potentials across $\Channels$ channels over $\OutputTime$ time steps, into the required output. It is common to further transform these continuous potentials into discrete events known as \textit{spikes}, representing neuronal action potentials, using a spike detection algorithm: $\ResponseSpikeMatrix = \SpikeDetection(\ResponseMatrix)$, where $\ResponseSpikeMatrix \in \Binary^{\Channels \times \OutputTime}$ is the binary matrix of detected spikes and $\SpikeDetection(\cdot)$ is the detection function \cite{Quiroga_2004_Unsupervised,pachitariu2016kilosort,chung2017fully,vargas2007automated}.

\subsection{Biological transformation}
\label{sec:biological_transform}

We model the BNN as a dynamical operator that maps the encoded stimulation matrix $\StimMatrix$ to the response matrix $\ResponseMatrix$.  
This mapping is non-stationary and depends on the temporal structure of the input, with parameters $\ParametersBNN$ evolving according to stimulation and response history, external feedback, as well as through intrinsic spontaneous processes. In this framework, the mapping $\BNN(\Cdot;\Cdot)$ can be interpreted as a qualitative change in the information contained in the input (in this case the stimulation matrix $\StimMatrix$ at time step $t$). We refer to qualitative change to a transformation that does not simply scale or linearly filter the input, but reorganizes its informational structure by mapping raw inputs into functionally meaningful representations.\footnote{E.g., continuous signals transformed into patterns that capture relational or dynamical features, revealing hidden structure not accessible from raw input.} The mapping $\BNN(\Cdot;\Cdot)$ therefore enables the BNN to extract and retain relevant features of the input, facilitating adaptivity of its internal state. This is expressed as:
%
% Short version
%
\vspace{-0.1cm}
\begin{equation}
    \ResponseMatrix = \BNN \left(\StimMatrix ; \ParametersBNN[t] \right), \quad \text{with} \;
    \ParametersBNN[t] = \BNNAdaptation \left( \AdaptationInput_{t-1} ; \ParametersBNN[t-1]\right),
\label{eq:b_adaptation}
\end{equation}
where the parameters $\ParametersBNN[t]$ change over time and the input $\AdaptationInput_{t-1}$ is drawn from $\{\ResponseMatrix[t-1], \Feedback(\Score[t-1] ; \ParametersFeedback), \varnothing\}$. We include $\varnothing$ to account for spontaneous biological adaptation in the absence of external interaction~\cite{sumi_biological_2023,Ciampi_2026_NeuroInspired}.

%
% Long version
%
% \begin{align}
%     \ResponseMatrix
% = 
%     \BNN \left(\StimMatrix ; \ParametersBNN[t] \right), 
% \end{align}
%
% while the set of parameters $\theta_{b,t}$ change over time as 
%
% \begin{align}
%     \ParametersBNN[t] = \BNNAdaptation \left( \AdaptationInput_{t-1} ; \ParametersBNN[t-1]\right),
% \label{eq:b_adaptation}
% \end{align}
%
% where the input $\AdaptationInput_{t-1}$ is drawn from $\{\ResponseMatrix[t-1], \varnothing\}$. The inclusion of $\varnothing$ accounts for possible biological adaptation in the absence of external interaction. 

The set of parameters $\ParametersBNN[t]$ serve as a mathematical proxy for the network's adaptive state, implicitly capturing underlying mechanisms such as specific plasticity rules, multiple timescales of adaptation, structural changes in connectivity, and biophysical processes, without explicitly modeling them. The recursive dependence \eqref{eq:b_adaptation} implies that the transformation implemented by the network is not fixed, but evolves as a function of the system's own activity. Thus, the response at time step $t$ depends not only on current and past input, but also on how previous outputs shaped the transformation itself. Systems with this property can be regarded as ``third-order'' information-processing systems~\cite{kagan_quantifiable_2026}. Here, the resulting transformation is non-invertible, reflecting reorganization and compression of information. Importantly, we treat encoding and decoding parameters as fixed for each learning phase, such that $\ParametersEncoder[t] = \ParametersEncoder[t-1]$ and $\ParametersDecoder[t] = \ParametersDecoder[t-1]$, so that adaptation is localized to the BNN.

\subsection{Feedback}

Finally, we define \textit{feedback} as a special form of encoding designed to purposefully drive BNN adaptation towards high-level objectives. Similarly, feedback is hardware agnostic and the techniques involved may range from purposefully crafted electrical stimulation to the dispensation of chemical mediators such as neurotransmitters \cite{robbins_goal-directed_2026, khajehnejad_dynamic_2025, jordan_open_2024, kagan_harnessing_2025}. Feedback is a stimulus delivered after the BNN responds to an initial stimulus, with the objective of modulating the mapping between that initial stimulus and the resulting biological state. Feedback can broadly be either 1)  \textit{reinforcing} ($\Reward$) to encourage and reinforce mappings that lead to favorable (from the perspective of the designer) task outcomes or, 2) \textit{plasticity inducing} ($\Punishment$) to encourage development of alternative mapping structures. 

\section{Empirical Evaluation}\label{section:empirical}

\begin{figure}[t]
    \centering
    \includegraphics[width=1.0\linewidth]{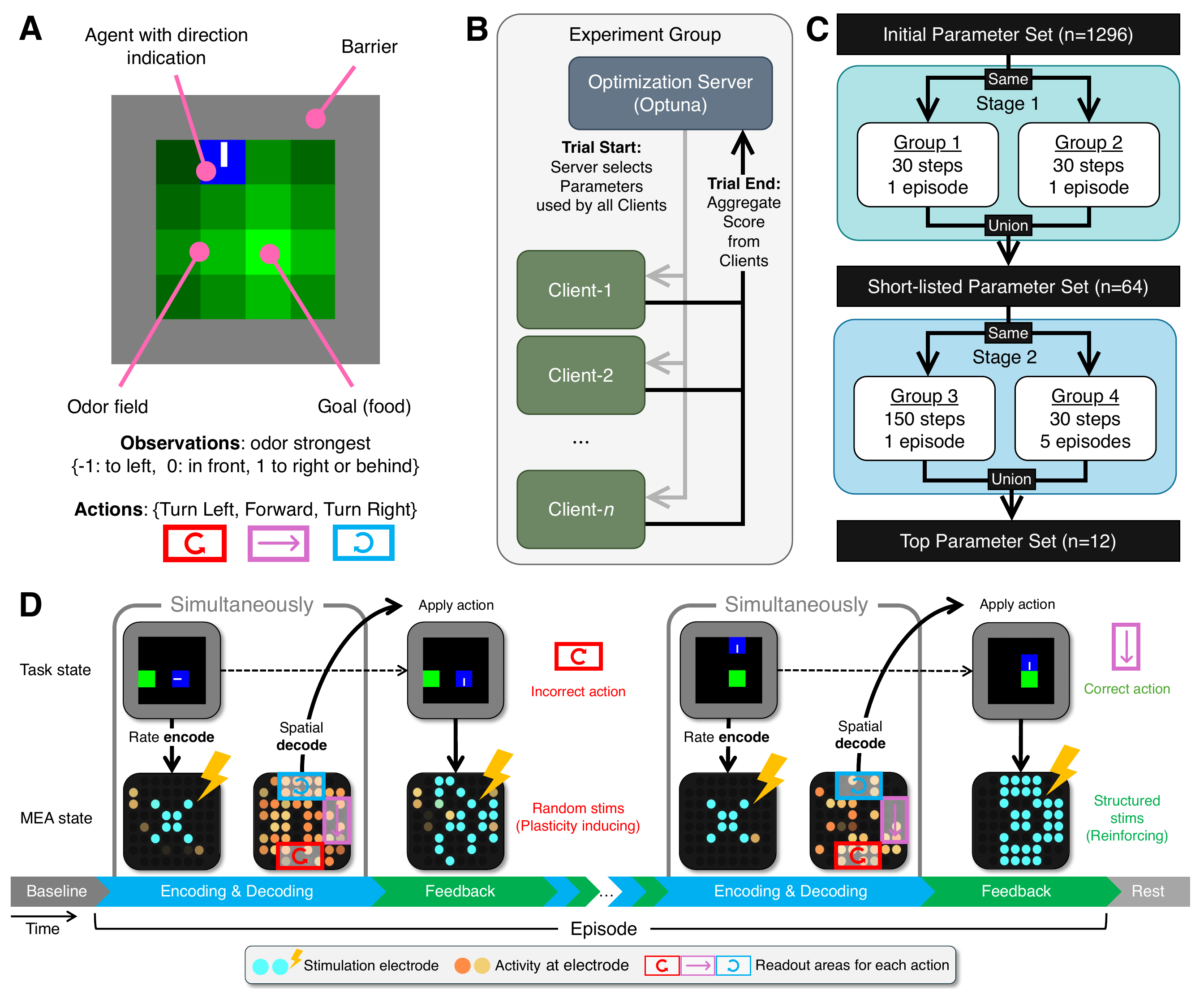}
    \caption{
    Experimental setup for framework evaluation. 
    \textbf{A}: Simulated goal-driven environment featuring an agent, food source, and odor gradient within a $6 \times 6$ gridworld. 
    \textbf{B}: Distributed optimization architecture using an Optuna server to dispatch parameters to parallel clients for aggregate trial scoring.
    \textbf{C}: Two-stage parameter screening pipeline: Stage 1 reduces 1,296 initial combinations to a shortlist ($n=64$), and Stage 2 identifies the final 12 optimal configurations.
    \textbf{D}: Closed-loop interaction cycle, illustrating task encoding, spatial decoding, and feedback types.
    }
    \label{fig:experiment}
    % \vspace{-15pt}
\end{figure}

\textbf{Task definition.}~ To demonstrate practical application of our framework, we create a simulated goal-driven navigation environment, which tasks an agent to navigate a barrier enclosed gridworld to reach a randomly spawned goal position referred to as \textit{food}, guided by a scalar sensor value referred to as \textit{odor} (see~\Cref{fig:experiment}A). In each step, the agent can take one of three actions: \texttt{move forward}, \texttt{turn left}, or \texttt{turn right}. Successful goal acquisition by the agent leads to a score increment which relocates the food, while barrier collisions decreases the score (\Cref{sec:supplementary_reward}). Task performance is evaluated on the total score acquired over a set number of steps across three modes: A) \texttt{30 steps over 1 episode}, B) \texttt{150 steps over 1 episode} and C) \texttt{30 steps over 5 episodes} spaced with 2 minutes rest. Episodes are reset using a common seed to ensure deterministic goal presentation.

\textbf{Experimental setup.}~ We implement a scaled distributed optimization platform consisting of an \textit{optimization server} using the Optuna hyperparameter optimization framework (HPO)~\cite{optuna_2019} and multiple connected Cortical Labs CL1 \textit{clients} via the Cortical Cloud service (see~\Cref{fig:experiment}B).\footnote{Accessible: https://corticallabs.com/cloud} Each CL1 encloses a single MEA comprising a unit of \textit{neuron culture}~\cite{kagan_cl1_2025}.\footnote{Power consumption for maintaining and interfacing biological substrates on the CL1 is quantified in~\Cref{sec:supplementary_energy}.} To address biological stochasticity, we evaluate identical parameter sets across multiple simultaneous CL1 clients, using their aggregated score as a single representative outcome. 

\textbf{Parameter optimization and evaluation.}~ We focus optimization on encoding hyperparameters $\ParametersEncoder$, which also dictate delivery of feedback pulses, while using relatively simple and fixed decoding and feedback mechanisms, to isolate their specific contributions from the broader, highly-coupled system dynamics. Specifically, we evaluate 1,296 parameter combinations for $\ParametersEncoder$ (see~\Cref{tab:parameters}) derived from a prioritized subset of six high-impact encoding parameters, selected via domain expertise to ensure an experimentally tractable search space. The optimization proceeds in two sequential stages across four experimental groups: Stage 1 (Groups 1 and 2) explores identical parameter sets under differing biological and hardware (biophysical) configurations using mode A, while Stage 2 (Groups 3 and 4) refines top performing Stage 1 parameters under common biophysical setup to evaluate modes B and C respectively (see~\Cref{fig:experiment}C). To ensure the stability of Stage 1 parameters, we performed multiple experimental replicates ($4$x) for each configuration in Stage 2, which results in a total of 256 trials.

\textbf{Encoding.}~ The environment presents a scalar sensor value (\Cref{sec:supplementary_reward}) which indicates that the odor field is strongest: (1) to the right or behind, (-1) to the left or (0) in front of the agent. Using a simplified representation allows us to decouple the study of BNN encoding techniques from the complexities of environment translation, while ensuring that modern silicon-based agents can achieve high performance to provide a stable benchmark for comparison. 

We use rate encoding (\Cref{section:supplementary_rate_encoding}), which translates the sensor value to a sequence of stimulations spaced by a frequency (Hz) that is interpolated between a \texttt{minimum frequency} (Hz) and \texttt{maximum frequency} (Hz). The \texttt{stimulation amplitude} (\uAmp) and \texttt{pulse width} (\uSec) define the charge delivered by symmetrical biphasic rectangular pulses with a negative-polarity leading phase. Finally, we refer to a \textit{tick} as a discrete unit of time in which rate encoding is delivered, parameterized by \texttt{tick rate} (Hz). The rate at which the agent interacts with the environment is measured in the number of ticks that passes for each step of the environment (referred to as \texttt{ticks per step}). The real-time speed of agent-environment interactions is governed by \texttt{tick rate} $/$ \texttt{ticks per step} (see~\Cref{fig:experiment}D and~\Cref{section:supplementary_temporal_rollout}). A breakdown of parameter values screened is listed in \Cref{tab:parameters}.

\textbf{Decoding.}~ For all experiments, we use a \textit{count decoding} approach adapted from~\cite{kagan_vitro_2022}. During each interaction period, spikes identified by the CL1's on-board spike detector are aggregated within three distinct spatial regions of the micro-electrode array (see~\Cref{section:supplementary_mea_layout}), where each region corresponds to a specific action. Total spike counts are normalized against a baseline of spontaneous activity spanning the 60 interactions immediately preceding each episode. The agent then executes the action associated with the region exhibiting the highest relative spike density.

\textbf{Feedback.}~ We use approaches adapted from~\cite{kagan_vitro_2022} to deliver feedback after each agent-environment interaction, guided by the environment \textit{reward} (see~\Cref{sec:supplementary_reward}). When reward value $>0$, we deliver five structured bursts of 100 Hz pulses (80 ms duration) spaced across the feedback duration to both encoding and decoding regions to reinforce neural activity. To induce plasticity (reward $\leq 0$), we use random stimulation to the same regions, where each stimulation electrode has an activation probability of $p=1/3$ and temporal pattern generated by sampling inter-stimulation intervals derived from a 3-25 Hz frequency range. For both modes, stimulation lasts twice as long as the preceding interaction period, with the same regimen used across all experiments. While stimulation amplitude and pulse width are inherited from the encoding configuration, feedback procedure remains fixed. % across all experiments.

\textbf{Biological.}~ For Stage 1, we screen identical parameter sets under task mode A using two distinct yet standard biophysical setups. For Group 1 we use seven cortical/hippocampal (1:1 ratio) cultures within a Polydimethylsiloxane (PDMS) ring, and nine cortical/hippocampal (1:1 ratio) cultures on astrocytes in a monolayer for Group 2. For Stage 2, we evaluate task modes B and C using comparable setups. In both Groups 3 and 4, we use five cortical-only cultures in a monolayer with a consistent ratio of cultures with and without astrocytes. See~\Cref{sec:supplementary_cell_culture_cryo,sec:supplementary_chip_preparation} for details. Depending on culture activity, Groups 1, 3, and 4 utilize CL1 as a mini-incubator with manual daily media changes, while Group 2 employs the onboard perfusion pump for automated exchange.

\begin{importantbox}
\textbf{Importantly}, the encoding, decoding, and feedback modules are fixed, i.e., non-trainable and non-adaptive, during each trial.
Hence all observed behavior related changes are due to the adaptation of the biological cultures during the trial.
\end{importantbox}

\section{Results}\label{section:results}

\subsection{Identifying top parameters}\label{section:results_top_parameters}

\begin{table*}[b]
\scriptsize
\vspace{-15pt}
  \caption{Experimental parameter combinations across stages, categorized by: rate encoding (min/max frequency), stimulation delivery (amplitude, pulse width), and temporal gameplay characteristics (tick rate, ticks per step) (see~\Cref{section:empirical}). \textbf{Bold} indicates top-performing parameters after each stage.}
  \vspace{1em}
  \label{tab:parameters}
  \centering
  \begin{tabular}{lccc}
    \toprule
    Parameter & \multicolumn{3}{c}{Experiment Stage} \\
    \cmidrule(r){2-4}
     & Stage 1 (n=1296) & Stage 2 (n=64) & Top (n=12) \\
    \midrule
    Min Frequency (Hz)  & 2.0, 3.0, \textbf{4.0}, 5.0 & \textbf{4.0} & 4.0 \\
    Max Frequency (Hz)  & \textbf{40.0}, \textbf{60.0}, \textbf{80.0}, \textbf{100.0} & \textbf{40.0}, \textbf{60.0}, \textbf{80.0}, 100.0 & 40.0, 60.0, 80.0 \\

    \midrule
    Amplitude ($\mu$A)  & 1.0, \textbf{2.0}, \textbf{2.5} & 2.0, \textbf{2.5} & 2.5 \\
    Pulse Width ($\mu$s) & \textbf{40.0}, \textbf{80.0}, 160.0 & \textbf{40.0}, \textbf{80.0} & 40.0, 80.0 \\

    \midrule
    Tick Rate (Hz)      & \textbf{1.0}, \textbf{2.0}, 4.0 & \textbf{1.0}, \textbf{2.0} & 1.0, 2.0 \\
    Ticks per Step      & \textbf{2}, \textbf{4}, 8 & 2, \textbf{4} & 4 \\
    \bottomrule
  \end{tabular}
\end{table*}

\begin{figure}
    \centering
    \includegraphics[width=0.9\linewidth]{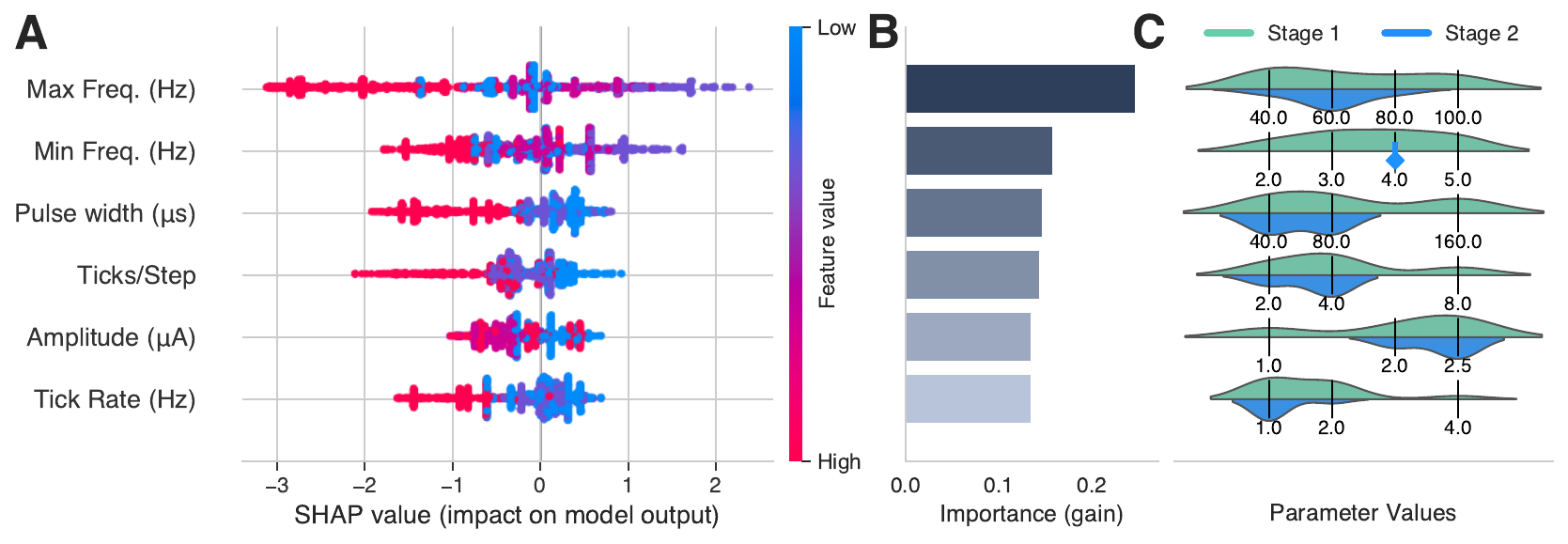}
    \caption{
    \textbf{A}: SHAP contributions to Stage 1 top 1\% performance (see~\Cref{section:results_top_parameters}).
    \textbf{B:} XGBoost feature importance in Stage 1. 
    \textbf{C:} Top 1\% parameter distributions across Stages 1 and 2.
    }
    \label{fig:parameter_importance}
    \vspace{-10pt}
\end{figure}

\Cref{tab:parameters} reports the range of parameters screened in each experimental stage as well as the top performing parameter values identified. To understand the importance of each parameter combination on overall performance, we fitted an XGBoost~\cite{Chen_2016_XGBoost} classifier with a binary objective to predict whether a parameter combination would yield a task score in the top 1\% (\texttt{class 1}) or otherwise (\texttt{class 0}). We use aggregate scores from Stage 1 where a minimum of 4 clients have responded to ensure each score is sufficiently representative of the parameter set's performance across multiple cultures. We then applied Shapley Additive Explanations (SHAP)~\cite{Lundberg_2017_unified,Lundberg_2020_local}, which quantifies the contribution of each feature toward the change in the model's expected output, enabling us to identify which encoding parameters most significantly drive high-performance outcomes. As shown in \Cref{fig:parameter_importance}, we find that: (1) maximum frequency for rate encoding is the strongest driver, favoring moderate values of 40-60 Hz, and (2) higher stimulation amplitude, shorter pulse width, and fewer ticks per step, corresponding to faster environmental interaction rates, all support improved performance.

Based on these insights, we reduced our initial pool of $n=1,296$ combinations to a shortlist of $n=64$ for Stage 2, from which we derived $n=12$ combinations exhibiting consistent improvement over baseline controls (\Cref{tab:parameters}). For  baselines, we use non-adaptive neuronal data via the CL-SDK Simulator\footnote{Accessible: https://docs.corticallabs.com/\#cl-sdk-simulator} comprising simulated random spikes (\textit{SDK Random}) and replayed spontaneous activity (\textit{Culture Baseline}). Specifically, for Stage 1, top performance is defined as trials exceeding the 99th percentile of \textit{SDK Random} scores across both Groups 1 and 2.

\subsection{BNNs learn to navigate}

\begin{figure}[t]
    \centering
    \includegraphics[width=1.0\linewidth]{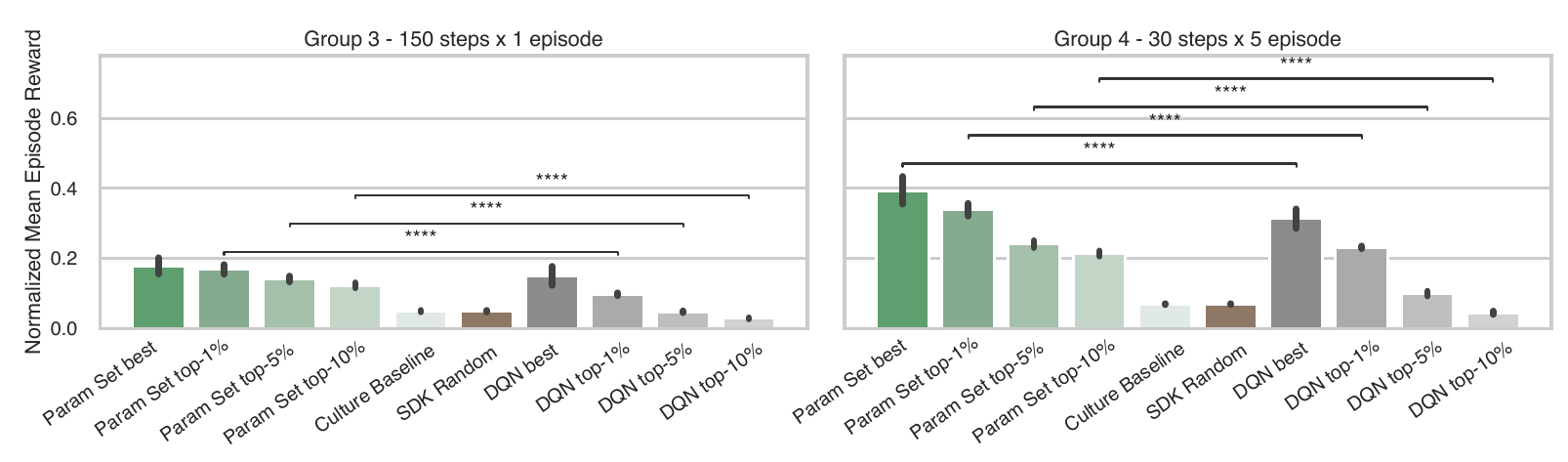}
    % \vspace{-20pt}
    \caption{Performance of BNN, DQN, and baseline agents in (left) Group 3 \texttt{150 steps $\times$ 1 episode} and (right) Group 4 \texttt{30 steps $\times$ 5 episodes}. Bars represent mean $\pm$ $95\%$ CI. Significance determined via Brunner-Munzel tests, where ****: $p < 10^{-4}$.}
    \label{fig:learning_comparison}
\end{figure}

In~\Cref{fig:learning_comparison}, we present Stage 2 results comparing Group 3 (a single 150-step episode) with Group 4 (five 30-step episodes separated by 2-minute rests). We report the normalized mean episode reward, which is the ratio of total reward to the maximum achievable reward. In both groups, optimized culture specific parameter sets demonstrated consistent performance gains over the baseline controls.

% \vspace{5pt}
\begin{figure}[t]
    \centering
    \includegraphics[width=1.0\linewidth]{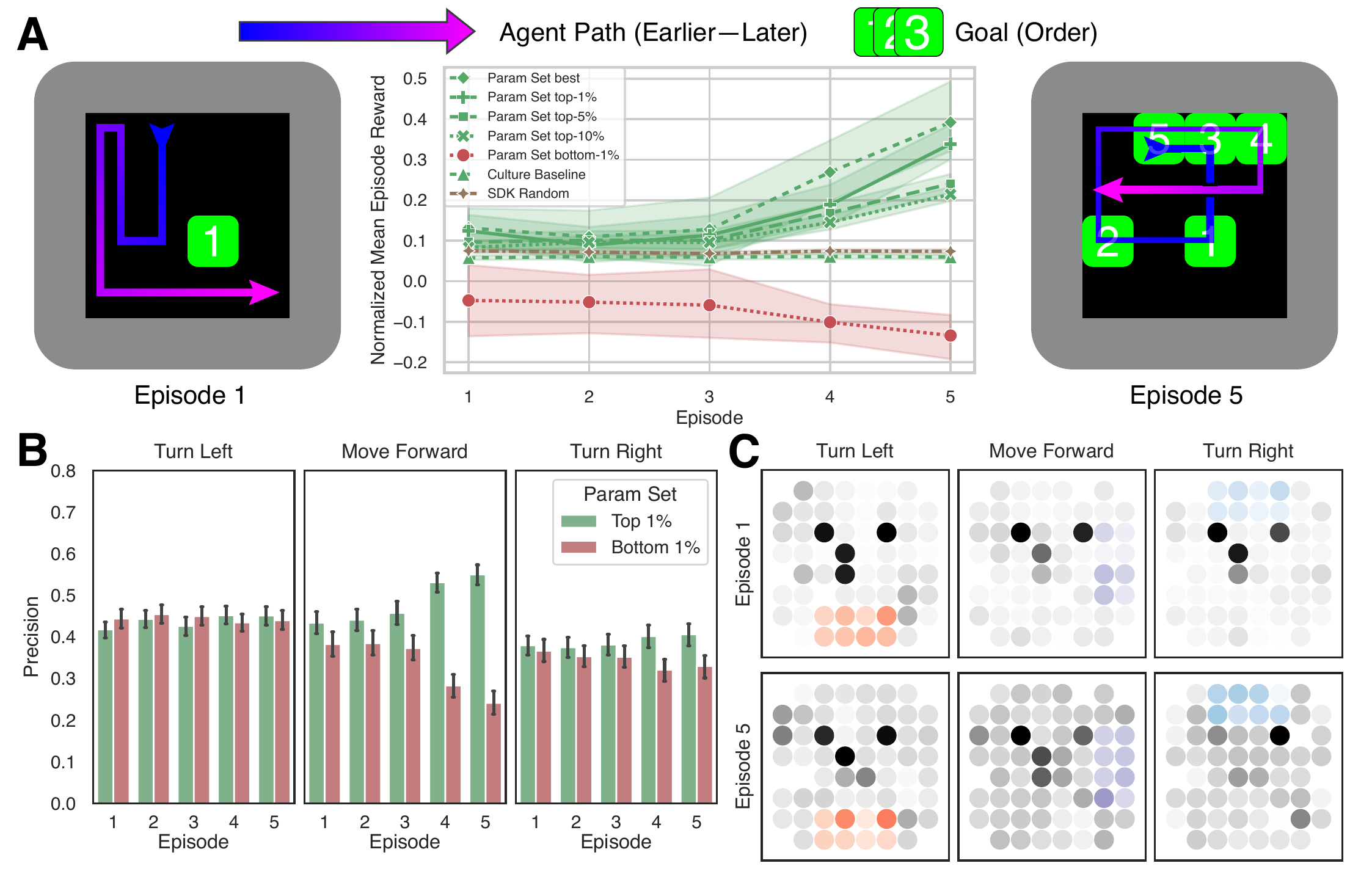}
    \caption{\textbf{A}: Group 4 learning progression, showing task performance (mean $\pm$ 95\% CI) across five episodes for various parameter sets and baselines, with examples of agent traces for episodes 1 and 5 (\textit{left} and \textit{right} respectively).
    \textbf{B}: Precision of decoded actions associated with reinforcing feedback (mean $\pm$ 95\% CI) for the top and bottom 1\% parameter sets in (A).
    \textbf{C}: Heat map showing average of relative spike counts across the MEA during episode 1 and 5 corresponding to examples in (A).}
    \vspace{-15pt}
    \label{fig:learning_episodes}
\end{figure}

We employ an optimized Deep-Q Network (DQN)~\cite{mnih2015human} as a representative of modern reinforcement learning to assess whether biological agents can acquire task-relevant behavior through learned strategies.
The DQN comparison puts the BNN performance into perspective, as comparisons between biological and \textit{classical} machine learning systems are inherently difficult. Rather than solely comparing maximum performance, it provides a reference for task difficulty and the qualitative degree of learning achieved by the BNN.
More details on the DQN setup can be found in Appendix~\ref{sec:supplementary_dqn}.
While the low-dimensional nature of this task does not necessitate high-capacity algorithms, we avoid rule-based approaches, such as heuristic decision trees or tabular methods, as they are insufficient for capturing the complex learning dynamics characteristic of biological and artificial neural networks.
Statistical analysis using Brunner-Munzel tests~\cite{brunner2000nonparametric} of the same top $x\%$ of parameters, which account for differences in sample sizes, distributions, and variances, shows that BNN agents achieve significantly higher performance than corresponding DQN groups (\Cref{fig:learning_comparison}) with $1.18$x and $1.25$x improvement for Group 3 and 4 respectively, and by $3.6$x and $5.6$x relative to culture baseline.

More interestingly, \Cref{fig:learning_comparison} shows that BNN agents achieve higher performance when learning is distributed across multiple episodes separated by rest periods, compared to a single continuous episode with equivalent number of interactions. In~\Cref{fig:learning_episodes}A, we observe a clear learning progression starting from episode three, with peak performance achieved by using optimal parameter sets for each BNN agent. Examples of behavioral traces for episode one and episode five (\textit{left} and \textit{right}, respectively) show a distinct shift toward goal-directed behavior. 
We find that this learning is characterized by a more strategic use of the \texttt{move forward} action, deploying it at the right moments to enable goal acquisition (\Cref{fig:learning_episodes}B).
This is supported by a distinct shift in network activity (\Cref{fig:learning_episodes}C) with initial activity primarily evoked near the encoding electrodes becoming more diverse and distributed across the decoding regions of the network by episode five. 

%
% This version places caption of Fig 6 to the right
%
\begin{figure}[b]
    \vspace{-15pt}
    \centering
    \begin{minipage}[c]{0.42\linewidth}
        \centering
        \includegraphics[width=\linewidth]{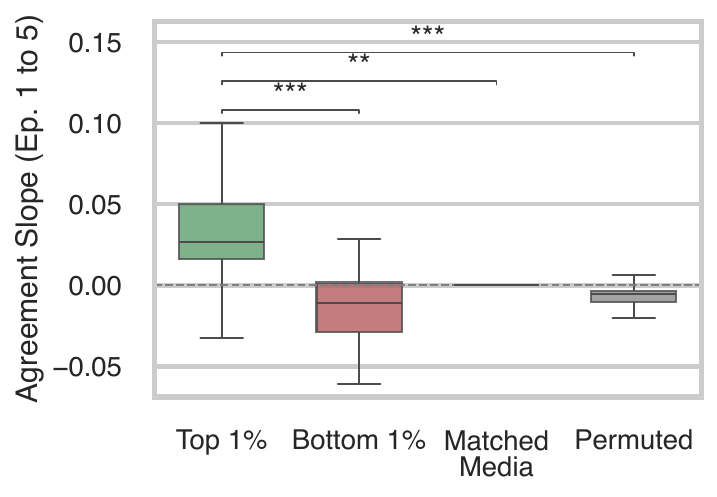}
    \end{minipage}
    \hfill
    \begin{minipage}[c]{0.54\linewidth}
        \caption{
        Agreement slope between observations and actions across Episodes 1 to 5 for the Top 1\% and Bottom 1\% configurations in Group 4, including Matched Media and Permuted ablation controls. Positive slope indicates the emergence of task-aligned learning.
        }
        \label{fig:slope_comparison}
    \end{minipage}
\end{figure}

%\vspace{5pt}
\subsection{Disentangling biological contribution} % or similar

%
% This version wraps Fig 6 and places the fig on the right of text
%
% Square wrapping use 18
% Full vertical wrapping use 26
% \begin{wrapfigure}[18]{R}{0.38\textwidth}
%     \vspace{-16pt}
%     \centering
%     \includegraphics[width=\linewidth]{figs/fig_6_slope_comparison_plot_paper.pdf}
%     \caption{
%     Agreement slope between observations and actions across Episodes 1 to 5 for the Top 1\% and Bottom 1\% configurations in Group 4, including Matched Media and Permuted ablation controls. Positive slope indicates the emergence of task-aligned learning.
%     }
%     \label{fig:slope_comparison}
% \end{wrapfigure}

Our task definition requires the BNN to adapt to fulfill a spatial gap between encoded observations and decoded actions. To determine whether the observed learning is attributable to biological processes, we test biological contribution through a formal hypothesis test (refer to~\Cref{sec:supplementary_biolearning} for full details and results). We evaluate: $H_1$: (Emergence), where the BNN adapts to bridge the spatial gap so the agreement between observations and actions increases over time, and $H_0$: (Null), where decoded action remains independent of the observation.

We apply this test to the Top and Bottom 1\% performing configurations in the Group 4 experiments. We also introduce two additional ablation controls, Permuted and Matched Media, to isolate the role of the BNN by permuting spatial decode mapping and using media without cultured neurons, respectively (\Cref{fig:slope_comparison}). The results support $H_1$ and contradict $H_0$, as the top-performing configurations demonstrate a significant positive agreement slope while the ablation controls exhibit near-zero agreement. These findings indicate that learning performance is both biologically driven and spatially dependent. Furthermore, the significant difference in learning rates between the Top and Bottom groups ($p < 0.001$) suggests that the emergence of biological learning is highly sensitive to hyperparameter selection, with the negative slope observed in the Bottom group serving as evidence of non-trivial, though non-task-aligned, biological adaptation.

% \vspace{-5pt}
\section{Limitations}
\label{section:limitations}
\vspace{-7pt}
\textbf{Search space constraints:} To ensure experimentally tractable parameter optimization, we restricted our search to a prioritized subset of encoding parameters within predefined ranges based on domain expertise. While this approach enabled an unprecedentedly large-scale screening of 1,296 combinations, it remains possible that optimal configurations reside beyond these initial boundaries. Nevertheless, the identified parameter-importance trends identified here provide a foundation for future work by guiding the expansion of the search space into broader encoding regimes or additional framework components, such as decoding and feedback.

\textbf{Temporal dependencies:} To prevent systematic bias, we employ a non-monotonic parameter delivery schedule across trials to avoid confounding of parameter values with trial progression. However, since the rest period between trials (2 minutes) is identical to the rest period between episodes in Group 4, carry-over effects from previous stimulation regimes cannot be ruled out. This uncertainty raises a compelling question regarding whether the performance lag observed in Group 4, where learning gains emerge only after the third episode, represents the emergence of longer-term adaptive processes. Although we refrain from using the term \textit{memory} as it is not yet formally defined within our framework, investigating these temporal couplings offers an exciting direction for future work to characterize the specific timescales required for retention and resetting of biological states.

\textbf{Task complexity and generalizability:} Our evaluation utilized a low-dimensional, scalar sensory input within a controlled gridworld environment as an initial proof of concept. While it remains unknown whether the integrated neurocomputational pipeline, comprising encoding, decoding, and feedback components, can generalize to high-dimensional, continuous sensory streams, our results provide a baseline for future comparative studies. This enables more purpose-driven interface strategies and their systematic validation in increasingly complex environments.

\textbf{Transferability of top parameters:} To reduce the possibility of culture-specific selection, identical parameter configurations were evaluated across multiple cultures, and their aggregated performance was used as score. We also included distinct biological culture configurations across the experimental groups. The resulting parameters were based on the average response of multiple cultures tested over multiple sessions. We applied these identified parameters to different cultures that were tested during different sessions. We found that learning was reproduced, suggesting that the top parameters have transferability across cultures. How far this transfer extends is a question for future work.

\textbf{Feedback-encoding overlap:} 
Since feedback is a specialized form of encoding, the two modules may not be fully independent and share a subset of stimulation parameters in our experiments. This design choice prevents a fully decoupled optimization of each component. However, we used a (partially) shared parameter setting to ensure that the observed effects reflect general stimulation properties rather than module-specific tuning. This provides a consistent baseline that can inform future work aimed at disentangling and optimizing stimulation protocols for each component individually.

\section{Conclusion}
\label{section:conclusion}
\vspace{-7pt}

We introduced \textit{Embodied Neurocomputation} as a systems-level framework for interfacing biological neuronal networks (BNNs) with digital environments through encoding, biological transformation, decoding, and feedback. We operationalized this framework as a coupled optimization problem and validated it through a large-scale evaluation of approximately 1,300 encoding configurations across 26 neural cultures, spanning $\sim$4,000 hours of real-time closed-loop agent-environment interactions.

Our results show that BNNs can support goal-oriented behavior when embedded in a sensorimotor loop. 
A small subset of encoding regimes produced consistent learning and significantly outperformed non-adaptive culture baselines and silicon DQN agents trained with the same number of interactions. 
These findings indicate that the computational potential of living neural substrates depends not only on the biological network itself, but also on the structure of the bio-silicon interface.

To validate the approach in this study the search space was necessarily constrained, with a low-dimensional sensory input, and the temporal dependencies underlying multi-episode improvement require further characterization. Future work should extend optimization to decoding and feedback, evaluate richer sensory environments, and directly probe the timescales and mechanisms of biological adaptation. By moving beyond heuristic stimulation protocols toward systematic interface optimization, this work provides a foundation for scalable, efficient, and adaptive bio-silicon computation, with long-term implications for low-power intelligent systems and hybrid biological-digital architectures.

\newpage
\bibliographystyle{unsrtnat}
\bibliography{references}

\newpage
\appendix
\section{Supplementary Materials}

\subsection{Cell Culture and Cryopreservation}\label{sec:supplementary_cell_culture_cryo}
hiPSC lines (CLV3.1, CLV4.4) were maintained under standard pluripotency conditions and differentiated into cortical neurons via a tri-inhibition protocol combining dual-SMAD inhibition with WNT and later MEK inhibition \cite{rosebrock_enhanced_2022}, or into hippocampal neurons (yielding DG, CA3, and CA1 subtypes, \cite{abu-bonsrah_novel_2026}). Cortical and hippocampal cultures were cryopreserved at DIV 30 and DIV 40, respectively, in CryoStor CS10 at $3 \times 10^{6}$ cells/mL using controlled-rate freezing and stored in liquid nitrogen until use. On the day of plating, vials were rapidly thawed at 37\Celcius, washed in plating medium (CTX medium \cite{abu-bonsrah_novel_2026} + chroman-1 50 nM), and assessed for viability by Trypan blue exclusion; only preparations exceeding $90\%$ viability were used.

\subsection{MEA Chip Preparation, Co-Culture Assembly, and Maturation}\label{sec:supplementary_chip_preparation}
Sixty-electrode MEA chips (Multi Channel Systems) were coated with poly-L-ornithine (0.1 mg/mL, Merck) followed by human laminin-521 (10 \ug /mL, StemCell Technologies) to promote neuronal adhesion. For cultures with primary human cortical astrocytes (ScienCell; passage +1 post-thaw), these were seeded slowly on to the electrode area in 5 \uL plating medium (as above) at 70,000 cells/chip into 200 \uL plating medium and allowed to form a confluent monolayer over up to 3 days before neuronal addition. For cultures without astrocytes, a Polydimethylsiloxane (PDMS) ring was placed around the electrode area and allowed to dry overnight. In co-cultures, thawed cortical and hippocampal neurons were co-seeded at 100,000 cells each (1:1 ratio; 200,000 total) either onto the astrocyte layer, or directly into the coated surface inside the PDMS ring, in 5 \uL plating medium, prior to flooding with 1 mL extra plating media after 1h. In cortical neuron-only cultures, the hippocampal component was excluded from the plating mixture. After 7 days, AraC 2 \uM was added for a single day to inhibit proliferation, and media was changed to BrainPhys + SM1 supplement (1X) + Pen Strep (1X) + Glutamax (1X) + BDNF (10 ng/mL) + dbcAMP (250 \uM) + ascorbic acid (200 \uM) thereafter. Cultures were maintained at 37°C, $5\%$ $\text{CO}_{2}$ with $50\%$ medium changes 3 times every week, and matured for 60 days post-seeding prior to electrophysiological and information-processing assessment.

\subsection{Micro-Electrode Array (MEA) Layout}\label{section:supplementary_mea_layout}

\begin{figure}[h]
    \centering
    \includegraphics[width=1.0\linewidth]{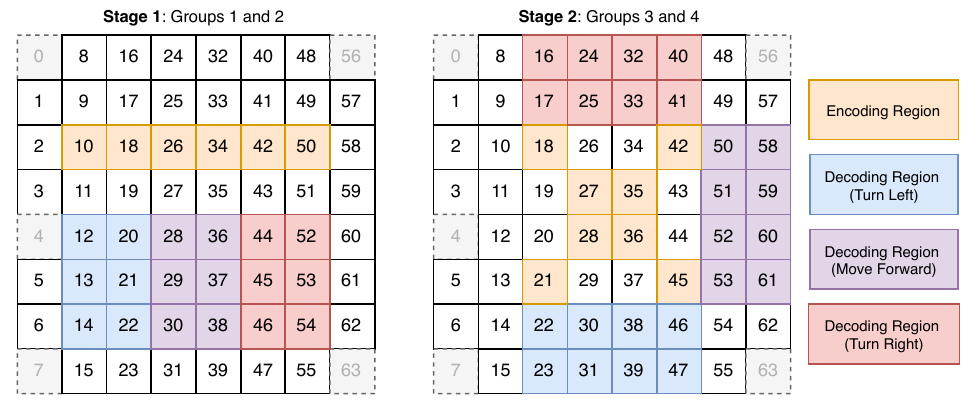}
    \caption{
    Spatial layouts used for encoding and decoding on the MEA. \textbf{Stage 1 Layout (left)}: A configuration placing encoding and decoding regions on opposite sides of the chip, which we find to  be limiting due to spatial heterogeneity in neuronal coverage. \textbf{Stage 2 Layout (right)}: An improved design positioning the encoding region centrally and equidistant from spatially separated decoding regions to mitigate spatial bias. In both configurations, we use the same number of encoding and decoding electrodes.
    }
    \label{fig:sup_mea_layout}
\end{figure}

\clearpage
\subsection{Rate Encoding}\label{section:supplementary_rate_encoding}

Rate encoding~\cite{Hua_2025_Microelectrode,kagan_vitro_2022,Adrian_1926_impulsesb,Adrian_1926_impulses,Adrian_1926_impulsesa} is a technique that communicates information to biological neurons by modulating the frequency of electrical stimulation. We implement rate encoding by interpolating scalar sensor information ($\ScalarInput$) between user-defined minimum ($\FrequencyMin$) and maximum ($\FrequencyMax$) frequencies, such that:

\begin{equation}
    \Frequency(\ScalarInput) = \FrequencyMin + (\FrequencyMax - \FrequencyMin) \left[ \frac{\ScalarInput - \ScalarInputMin}{\ScalarInputMax - \ScalarInputMin} \right].
\end{equation}

\begin{figure}[h]
    \centering
    \includegraphics[width=1.0\linewidth]{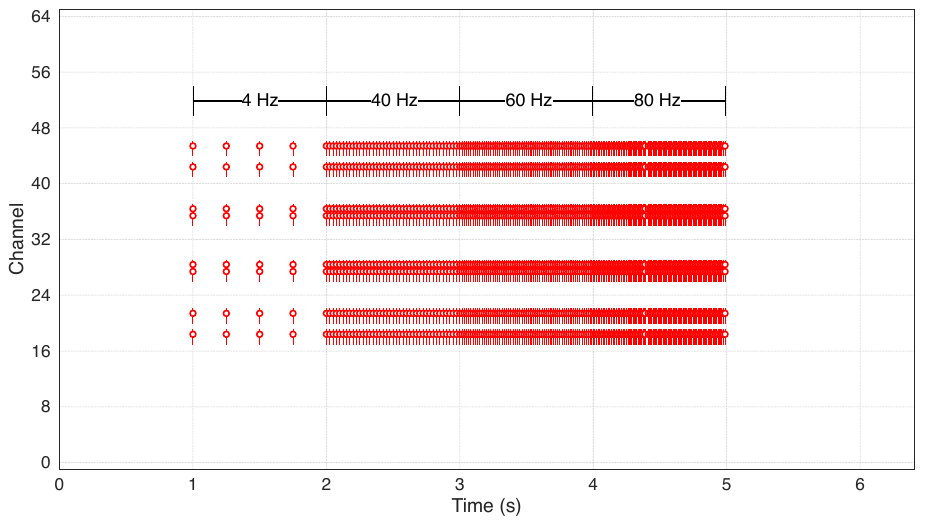}
    \caption{
    Raster plot showing four examples of rate encoding on the encoding channels being delivered at 4 Hz, 40 Hz, 60 Hz, and 80 Hz for 1 second each, with stimulation pulses marked in red.
    }
    \label{fig:sup_rate_encoding}
\end{figure}

At each environment interaction, we deliver rate encoding simultaneously (see~\Cref{fig:sup_rate_encoding}) to all electrodes within the defined encoding region (\Cref{fig:sup_mea_layout}). The duration of encoding is defined by the encoding parameters \texttt{tick rate} and \texttt{ticks per step}.

\clearpage
\subsection{Temporal Rollout}\label{section:supplementary_temporal_rollout}

\begin{figure}[h]
    \centering
    \includegraphics[width=1.0\linewidth]{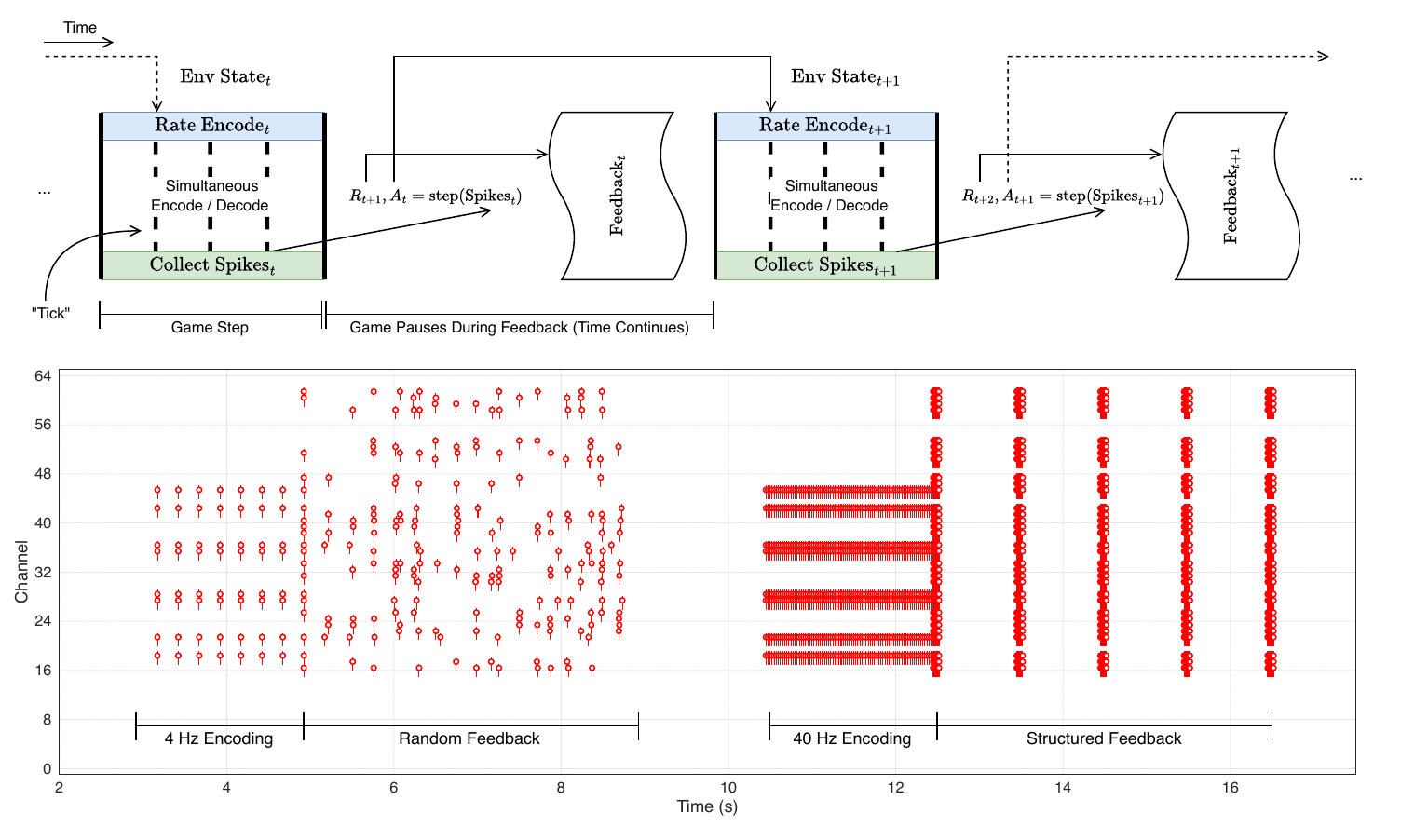}
    \caption{
    Schematic of temporal unfolding of environment state encoding and decoding (above) with an example raster plot marking stimulation pulses in red that correspond to two environment interactions: (1) 4 Hz encoding followed by random feedback, and (2) 40 Hz encoding followed by structured feedback.
    }
    \label{fig:sup_temporal_rollout}
\end{figure}

Schematic of temporal unfolding of environment state encoding and decoding is shown in~\Cref{fig:sup_temporal_rollout}. At each game step $t$, sensor information is encoded and delivered as electrical stimulation to the biological neural networks (BNNs) via rate encoding while simultaneously collecting spikes detected by the CL1 device. The $\text{step}(\Cdot)$ function advances the environment to the next timestep by decoding an action $A_{t}$ from the spike events using count decoding and moving the agent. A reward value is returned for this interaction $R_{t+1}$ which initiates suitable Feedback stimulation.

\subsection{Reward and Sensing Dynamics} \label{sec:supplementary_reward}

The reward function $R_t$ at timestep $t$ (see~\Cref{fig:sup_temporal_rollout}) is defined by the agent's interaction with the environment components, specifically the goal (\textit{food}) and obstacle boundaries. The reward signal is composed of three distinct terms: 
(1) a terminal-style reward of $+2$ upon successful food acquisition; 
(2) a collision penalty of $-0.2$ when the agent strikes an obstacle or boundary; and 
(3) a continuous term representing the relative change in sensed odor intensity otherwise.

The sensing mechanism relies on an \textit{odor sensor} $\mathcal{S} \in \{-1, 0, +1\}$, that tracks relative changes in an \textit{odor field}, with intensity that decays exponentially with the distance $d$ from the food source, formulated as $I(d) \propto e^{-\lambda d}$, where $\lambda$ is the decay constant. The gradient of the field is discretized into three directional states that indicates the highest odor intensity relative to the agent's heading: 
(i) left $[-1]$, 
(ii) directly in front of the agent $[0]$,
(iii) and, right or behind $[+1]$.

\newpage
\subsection{Deep-Q Agent}\label{sec:supplementary_dqn}

For the \textit{DQN Agent}, we used a hidden layer size of $8$ with \texttt{ReLU} non-linearity and an online learning setup matching the BNN setup, i.e., learning from each experience  once, with the same number of total environment interactions as the BNN agents.
We optimized the parameters \texttt{learning rate}, \texttt{target network update frequency}, \texttt{train frequency} indicating the number of steps before training is triggered, \texttt{final exploration} the final $\epsilon$ for the exploration, and \texttt{exploration horizon} the ratio of training steps for annealing $\epsilon$. In total, we tested around ~$9,500$ parameter configurations to optimize these settings.

We report cut-off-early scores, for both, the reported DQN as well as the BNN agents, to compare their performance and learning dynamics given the same environment interaction budget. The parameters of the DQN are optimized for this learning horizon for a fair comparison. We ran additional DQN experiments with an increased budget of up to 1,000 episodes (200x times the original budget), using the original optimized parameters as well as using new optimized parameters for this setting. The DQN agents converge to a maximum normalized score around $0.75$. Note that the BNN agents also seem to be not converged yet (see Fig.~\ref{fig:learning_episodes}A). We also evaluated the number of episodes the DQN agents require to match the performance of the BNN agents after 5 episodes: all tested and optimized parameter settings for the DQN agents required at least four times more episodes as the BNN agents to equal their score.

\subsection{Biological learning as a testable hypothesis} \label{sec:supplementary_biolearning}

We model the task as the transformation $y_t = d(b(e(x_t,\theta_e), \theta_b),\theta_d)$ (Equation~\ref{eq:model_composition}), where:
\begin{itemize}
\item  $e(\cdot, \theta_e)$ is a fixed, spatially uniform, rate encoder that interpolates ordinal observation $x_t$ into a scalar stimulation frequency,
\item  $d(\cdot, \theta_d)$ is a fixed, count based decoder over three spatially-distinct regions representing actions $y_t$, and
\item  $b(\cdot, \theta_b)$ is the adaptive biological neural network (BNN).
\end{itemize}

Since encoding is centralized and decoding is spatially distributed, the BNN must perform a non-linear spatial reorganization to map a central frequency based stimulus to one of the three equidistant periphery regions. More specifically, encoding simultaneously activates all electrodes within the encoding region with same stimuli intensity (yellow in Figure~\ref{fig:sup_mea_layout}) and placement is equidistant and distinct from the decoder regions, creating a spatial gap.

This spatial gap ensures that any potential correlation between the encoded signal and the target action must be mediated by the BNN to bridge the physical distance between the input and output electrode regions. Since the parameters $\theta_e$ and $\theta_d$ are held fixed for all episodes within a trial, any emergence of task-relevant behavior must be attributable to changes in biology $b(\cdot, \theta_{b,t})$, which must transform the signal to satisfy this spatial requirement.

We, therefore, treat biological learning as a testable hypothesis through the following formulations:

\begin{itemize}
    \item $H_1$ \textbf{(Emergence):} The BNN adapt such that the \textit{agreement} between the decoded action $y$ and the observation $x$ exceeds chance and increases over time. This is suggestive of the emergence of biological learning to transform a central temporal stimulus into a specific spatial response. Formally, the mapping $x \xrightarrow{b} y$ is valid if and only if $b(., \theta_b)$ induces a non-trivial spatial reorganization of the input signal.
    \item $H_0$ \textbf{(Null Hypothesis):} The decoded action $y$ remains independent of the observation $x$. This is representative of a \textit{passive relay} scenario, where the biological network fails to reorganize, and the task performance remains at chance levels because the fixed encoder cannot bridge the spatial gap to the decoder. Formally, we refer to \textit{passive relay} where the mapping $x \rightarrow y$ remains essentially a stochastic or identity-mismatched function, failing to establish a correlation between input $x$ and correct action $y$ despite BNN adaptation.
\end{itemize}

We define agreement as the interaction accuracy, $P(y_t=x_t)$, representing the probability of the decoded action matching the correct action given the observation per time step $t$. We use the slope of agreement, which is the change in accuracy across episodes, to quantify a rate of learning and calculate the difference in slope between the Top percentile and each comparison group for hypothesis testing. We compute a 95$\%$ confidence interval by resampling the trials 2,000 times and determine significance ($p$-value) via a permutation test with 2,000 shuffles.

\begin{table}[h]
    %\footnotesize
    \caption{Table reports agreement for the Group 4 experiments, chance agreement is $0.33$.}
    \label{tab:agreement_learning}
    \centering
    \begin{tabular}{
        ll
        S[table-format=1.4]
        S[table-format=1.4]
        S[table-format=1.4]
        S[table-format=1.4]
        S[table-format=1.4]
        S[table-format=+1.4]
    }
        \toprule
        Percentile
        & Comparison Grp.
        & \multicolumn{1}{c}{Ep. 1}
        & \multicolumn{1}{c}{Ep. 2}
        & \multicolumn{1}{c}{Ep. 3}
        & \multicolumn{1}{c}{Ep. 4}
        & \multicolumn{1}{c}{Ep. 5}
        & \multicolumn{1}{c}{Slope/Ep.} \\
        \midrule

        $1\%$
        & Top
        & {\bfseries 0.3929}
        & 0.3265
        & {\bfseries 0.3648}
        & {\bfseries 0.3903}
        & {\bfseries 0.5281}
        & {\bfseries +0.0334} \\

        $1\%$
        & Bottom
        & 0.2143
        & 0.1920
        & 0.1964
        & 0.1801
        & 0.1592
        & -0.0122 \\

        $1\%$
        & Matched Media
        & 0.0657
        & 0.0979
        & 0.0936
        & 0.0893
        & 0.0929
        & +0.0046 \\

        $1\%$
        & Permuted
        & 0.3214
        & {\bfseries 0.3347}
        & 0.3270
        & 0.3219
        & 0.2944
        & -0.0067 \\

        \midrule

        $5\%$
        & Top
        & {\bfseries 0.3668}
        & {\bfseries 0.3577}
        & {\bfseries 0.3662}
        & {\bfseries 0.3980}
        & {\bfseries 0.4365}
        & {\bfseries +0.0180} \\

        $5\%$
        & Bottom
        & 0.2372
        & 0.2357
        & 0.2255
        & 0.2036
        & 0.2082
        & -0.0090 \\

        $5\%$
        & Matched Media
        & 0.0665
        & 0.1117
        & 0.0947
        & 0.1062
        & 0.0962
        & +0.0054 \\

        $5\%$
        & Permuted
        & 0.3266
        & 0.3285
        & 0.3268
        & 0.3204
        & 0.3127
        & -0.0036 \\

        \midrule

        $10\%$
        & Top
        & {\bfseries 0.3545}
        & {\bfseries 0.3525}
        & {\bfseries 0.3569}
        & {\bfseries 0.3818}
        & {\bfseries 0.4196}
        & {\bfseries +0.0159} \\

        $10\%$
        & Bottom
        & 0.2774
        & 0.2594
        & 0.2685
        & 0.2409
        & 0.2375
        & -0.0098 \\

        $10\%$
        & Matched Media
        & 0.0664
        & 0.1079
        & 0.0902
        & 0.1014
        & 0.0973
        & +0.0055 \\

        $10\%$
        & Permuted
        & 0.3291
        & 0.3295
        & 0.3286
        & 0.3236
        & 0.3161
        & -0.0032 \\

        \bottomrule
    \end{tabular}
\end{table}

\begin{table}[h]
    \caption{Statistical significance of the learning rate difference (Top vs. Comparison Group)}
    \label{tab:significance_learning}
    \centering
    \begin{tabular}{
        ll
        S[table-format=+1.4]
        S[table-format=+1.4]
        @{[}S[table-format=+1.4]@{,\ }S[table-format=+1.4]@{]}
        S[table-format=1.4]
    }
        \toprule
        Percentile
        & Comparison Grp.
        & \multicolumn{1}{c}{Median slope}
        & \multicolumn{1}{c}{Mean slope}
        & \multicolumn{2}{c}{IQR}
        & \multicolumn{1}{c}{Sig. vs. Top} \\
        \midrule

        $1\%$
        & Top
        & {\bfseries +0.0268}
        & {\bfseries +0.0334}
        & +0.0161
        & +0.0500
        & {-} \\

        $1\%$
        & Bottom
        & -0.0107
        & -0.0122
        & -0.0286
        & +0.0018
        & {\bfseries 0.0005} \\

        $1\%$
        & Matched Media
        & +0.0000
        & +0.0046
        & +0.0000
        & +0.0000
        & {\bfseries 0.0030} \\

        $1\%$
        & Permuted
        & -0.0054
        & -0.0067
        & -0.0100
        & -0.0032
        & {\bfseries 0.0005} \\

        \midrule

        $5\%$
        & Top
        & {\bfseries +0.0179}
        & {\bfseries +0.0180}
        & +0.0036
        & +0.0339
        & {-} \\

        $5\%$
        & Bottom
        & -0.0089
        & -0.0090
        & -0.0214
        & +0.0036
        & {\bfseries 0.0005} \\

        $5\%$
        & Matched Media
        & +0.0000
        & +0.0054
        & +0.0000
        & +0.0143
        & {\bfseries 0.0015} \\

        $5\%$
        & Permuted
        & -0.0036
        & -0.0036
        & -0.0068
        & -0.0007
        & {\bfseries 0.0005} \\

        \midrule

        $10\%$
        & Top
        & {\bfseries +0.0143}
        & {\bfseries +0.0159}
        & +0.0036
        & +0.0286
        & {-} \\

        $10\%$
        & Bottom
        & -0.0107
        & -0.0098
        & -0.0214
        & +0.0036
        & {\bfseries 0.0005} \\

        $10\%$
        & Matched Media
        & +0.0000
        & +0.0055
        & +0.0000
        & +0.0134
        & {\bfseries 0.0010} \\

        $10\%$
        & Permuted
        & -0.0029
        & -0.0032
        & -0.0057
        & -0.0007
        & {\bfseries 0.0005} \\

        \bottomrule
    \end{tabular}
\end{table}

\begin{table}[ht]
    %\footnotesize
    \caption{Full results from the decoder permutation. Permuted 1 and 2 maps actions as \texttt{[Left, Forward, Right]} and \texttt{[Right, Forward, Left]}, respectively. Since the emergent learning is mainly attributable to the Forward action (Fig. \ref{fig:learning_episodes}B), a lower positive learning slope is retained.}
    \label{tab:full_permutation}
    \centering
    \begin{tabular}{
        ll
        S[table-format=1.4]
        S[table-format=1.4]
        S[table-format=1.4]
        S[table-format=1.4]
        S[table-format=1.4]
        S[table-format=+1.4]
    }
        \toprule
        Perc.
        & Decode
        & \multicolumn{1}{c}{Ep. 1}
        & \multicolumn{1}{c}{Ep. 2}
        & \multicolumn{1}{c}{Ep. 3}
        & \multicolumn{1}{c}{Ep. 4}
        & \multicolumn{1}{c}{Ep. 5}
        & \multicolumn{1}{c}{Slope/Ep.} \\
        \midrule

        $1\%$
        & Permuted 1 (true)
        & {\bfseries 0.3929}
        & 0.3265
        & {\bfseries 0.3648}
        & {\bfseries 0.3903}
        & {\bfseries 0.5281}
        & {\bfseries +0.0334} \\

        $1\%$
        & Permuted 2
        & 0.3342
        & 0.3010
        & 0.3316
        & 0.3699
        & 0.4082
        & +0.0217 \\

        $1\%$
        & Permuted 3
        & 0.3597
        & 0.3801
        & 0.3418
        & 0.3189
        & 0.3469
        & -0.0087 \\

        $1\%$
        & Permuted 4
        & 0.3061
        & 0.3189
        & 0.3265
        & 0.3112
        & 0.2449
        & -0.0130 \\

        $1\%$
        & Permuted 5
        & 0.3138
        & 0.3648
        & 0.3597
        & 0.3163
        & 0.2730
        & -0.0130 \\

        $1\%$
        & Permuted 6
        & 0.2934
        & 0.3087
        & 0.2755
        & 0.2934
        & 0.1990
        & -0.0204 \\

        \midrule

        $5\%$
        & Permuted 1 (true)
        & {\bfseries 0.3668}
        & 0.3577
        & {\bfseries 0.3662}
        & {\bfseries 0.3980}
        & {\bfseries 0.4365}
        & {\bfseries +0.0180} \\

        $5\%$
        & Permuted 2
        & 0.3464
        & 0.3294
        & 0.3452
        & 0.3628
        & 0.3815
        & +0.0104 \\

        $5\%$
        & Permuted 3
        & 0.3611
        & 0.3549
        & 0.3577
        & 0.3424
        & 0.3407
        & -0.0053 \\

        $5\%$
        & Permuted 4
        & 0.2925
        & 0.3158
        & 0.2971
        & 0.2948
        & 0.2778
        & -0.0050 \\

        $5\%$
        & Permuted 5
        & 0.3430
        & 0.3588
        & 0.3566
        & 0.3407
        & 0.3277
        & -0.0049 \\

        $5\%$
        & Permuted 6
        & 0.2902
        & 0.2834
        & 0.2772
        & 0.2613
        & 0.2358
        & -0.0131 \\

        \midrule

        $10\%$
        & Permuted 1 (true)
        & {\bfseries 0.3545}
        & 0.3525
        & 0.3569
        & {\bfseries 0.3818}
        & {\bfseries 0.4196}
        & {\bfseries +0.0159} \\

        $10\%$
        & Permuted 2
        & 0.3447
        & 0.3452
        & 0.3569
        & 0.3664
        & 0.3795
        & +0.0091 \\

        $10\%$
        & Permuted 3
        & 0.3560
        & 0.3499
        & 0.3577
        & 0.3476
        & 0.3415
        & -0.0031 \\

        $10\%$
        & Permuted 4
        & 0.2994
        & 0.3049
        & 0.2854
        & 0.2860
        & 0.2790
        & -0.0060 \\

        $10\%$
        & Permuted 5
        & 0.3513
        & 0.3612
        & 0.3696
        & 0.3548
        & 0.3426
        & -0.0024 \\

        $10\%$
        & Permuted 6
        & 0.2941
        & 0.2863
        & 0.2735
        & 0.2634
        & 0.2378
        & -0.0136 \\

        \bottomrule
    \end{tabular}
\end{table}

\Cref{tab:agreement_learning,tab:significance_learning} report results stratified by percentile (1$\%$, 5$\%$, and 10$\%$). The comparison groups include the Top percentile, the Bottom percentile, as well as two ablation controls being Permuted group (combinatorial permutation of decoding regions\footnote{\label{fn:permuted_group}Permuted group involves 6 combinatorial permutations where Permuted 1 is the true mapping and Permuted 2 to 6 are shuffled. For brevity, we have reported the aggregate average in Tables~\ref{tab:agreement_learning} and \ref{tab:significance_learning}, while Table~\ref{tab:full_permutation} shows the full Permuted results.}, and the Matched Media group (no-cells control).

The data shown in Tables~\ref{tab:agreement_learning} and \ref{tab:significance_learning} support $H_1$ and are inconsistent with $H_0$. In the Top encoding configurations, the BNN acquires the required mapping across episodes, with agreement climbing from approximately 0.39 to 0.53. In Bottom configurations, agreement remains below chance and continues to drift lower. The learning-rate difference is significant at every threshold (permutation $p < 0.001$).

Comparison with the two controls demonstrates that the performance can be both attributable to the biology and is dependent on spatial-mapping. Media-only (No Cells) involves replaying the game on media without cultured neurons using parameter-matched configurations. This group shows flat agreement (slope $\approx 0$) and remains far below chance ($\sim$0.10), suggesting that the task cannot be solved by the silicon interface alone. Permute Decode\footref{fn:permuted_group} involves replaying the top recordings with the permuted decoding mapping. This collapses the learning trend to chance (slope $\approx -0.006$), demonstrating that the task performance depends on the correct spatial-mapping for decoding rather than a dominant encoder driven policy. 

Further, the marked divergence between Top and Bottom groups suggests that not only the biology is attributable to performance, configuration parameters can have a marked impact on whether the required spatial-mappings are developed. While the slope is negative for the Bottom group, it is important to note that the group starts from a significantly higher baseline ($\sim$0.21) than the Matched Media group ($\sim$0.10). We interpret the negative slope as evidence that the BNN is undergoing non-trivial (though non-task-aligned) biological adaptation, whereas the Matched Media group remains non-adaptive. As shown in~\Cref{tab:media_vs_bottom}, the difference in the learning rates (slopes) between these two groups is statistically significant.

% 

%
% Align decimals and sign
%

\begin{table}[h]
    \footnotesize
    \caption{Comparison of slope estimates between Media and Bottom across percentiles.}
    \label{tab:media_vs_bottom}
    \centering
    \begin{tabular}{
        ll
        S[table-format=+1.4]
        S[table-format=+1.4]
        S[table-format=+1.4]
        @{[}S[table-format=+1.4]@{,\ }S[table-format=+1.4]@{]}
        S[table-format=1.4]
    }
        \toprule
        Percentile
        & Comparison
        & \multicolumn{1}{c}{Slope (Media)}
        & \multicolumn{1}{c}{Slope (Bottom)}
        & \multicolumn{1}{c}{Difference}
        & \multicolumn{2}{c}{95\% CI}
        & \multicolumn{1}{c}{Perm. $p$} \\
        \midrule

        $1\%$
        & Media vs Bottom
        & +0.0046
        & -0.0122
        & +0.0168
        & +0.0052
        & +0.0285
        & {\bfseries 0.0055} \\

        $5\%$
        & Media vs Bottom
        & +0.0054
        & -0.0090
        & +0.0144
        & +0.0084
        & +0.0206
        & {\bfseries 0.0005} \\

        $10\%$
        & Media vs Bottom
        & +0.0055
        & -0.0098
        & +0.0154
        & +0.0112
        & +0.0198
        & {\bfseries 0.0005} \\

        \bottomrule
    \end{tabular}
\end{table}

\newpage
\subsection{Quantifying Energy Usage} \label{sec:supplementary_energy}

It has been estimated that, the human brain consumes approximately 10-20 W with processing capacity of around 1 exaFLOPS~\cite{Smirnova_2023_Organoid}. However, a direct comparison using standard silicon metrics like FLOPS per Watt is problematic for two primary reasons:

\begin{enumerate}
    \item Biological Neural Networks (BNNs) operate on non-von Neumann architectures where information processing relies on analogue, stochastic, and biochemical processes rather than discrete floating-point operations, and
    \item the precise mechanistic formulation of BNN computation remains an open question in neuroscience.
\end{enumerate}

Because BNNs do not perform classical arithmetic, they cannot be evaluated using floating-point metrics. We argue that our contribution, the Embodied Neurocomputation Framework, provides the necessary formal structure to allow for scalable and systematic studies using BNNs. As interaction techniques mature and the mechanisms through which BNNs process information become clearer, we expect that BNN processing capacities can be quantified. This will enable the development of fair and realistic energy benchmarks to compare biological and silicon-based architectures in the future.

Although we cannot perform a direct FLOPS comparison, we provide empirical data regarding the power consumption of the hardware supporting the biological substrate. Using 96,186 data points collected from 67 CL1 units over a two-day period, we quantified the power consumption required to maintain the biological interface. The results are summarized in~\Cref{tab:table_energy}.

\begin{table}[h]
    \vspace{-10pt}
    \caption{Power consumption required to maintain the biological interface.}
    \label{tab:table_energy}
    \centering
    \begin{tabular}{cc}
        Percentile & Power (W) \\
        \toprule
        1$\%$ & 10.6 \\
        5$\%$ & 17.5 \\
        10$\%$ & 18.0 \\
        50$\%$ & 22.7 \\
        90$\%$ & 23.8 \\
        95$\%$ & 25.2 \\
        99$\%$ & 25.5 \\
        \bottomrule
    \end{tabular}
\end{table}

This data represents hardware power consumption required for interaction and support of biological substrate under typical load within the Cortical Cloud service. In contrast, typical modern GPU enabled workstations may consume on average 579-641W~\cite{Wegmeth_2025_Green}. We believe this provides a practical foundation for future researchers to calculate the relative efficiency of BNNs against silicon-based systems as biological formalisms mature. Such comparisons must account for the fact that neuron computation capacity is not yet well characterized, which makes direct comparisons with arithmetic-based metrics such as FLOPS incompatible.

\end{document}